# Scale- free networks in cell biology

Réka Albert, Department of Physics and Huck Institutes of the Life Sciences, Pennsylvania State University

## Summary

A cell's behavior is a consequence of the complex interactions between its numerous constituents, such as DNA, RNA, proteins and small molecules. Cells use signaling pathways and regulatory mechanisms to coordinate multiple processes, allowing them to respond to and adapt to an ever-changing environment. The large number of components, the degree of interconnectivity and the complex control of cellular networks are becoming evident in the integrated genomic and proteomic analyses that are emerging. It is increasingly recognized that the understanding of properties that arise from whole-cell function require integrated, theoretical descriptions of the relationships between different cellular components. Recent theoretical advances allow us to describe cellular network structure with graph concepts, and have revealed organizational features shared with numerous non-biological networks. How do we quantitatively describe a network of hundreds or thousands of interacting components? Does the observed topology of cellular networks give us clues about their evolution? How does cellular networks' organization influence their function and dynamical responses? This article will review the recent advances in addressing these questions.

## Introduction

Genes and gene products interact on several level. At the genomic level, transcription factors can activate or inhibit the transcription of genes to give mRNAs. Since these transcription factors are themselves products of genes, the ultimate effect is that genes regulate each other's expression as part of gene regulatory networks. Similarly, proteins can participate in diverse post-translational interactions that lead to modified protein functions or to formation of protein complexes that have new roles; the totality of these processes is called a protein-protein interaction network. The biochemical reactions in the cellular metabolism can likewise be integrated into a metabolic network whose fluxes are regulated by enzymes catalyzing the metabolic reactions. In many cases these different levels of interactions are integrated - for example, when the presence of an external signal triggers a cascade of interactions that involves both biochemical reactions and transcriptional regulation.

A system of elements that interact or regulate each other can be represented by a mathematical object called graph (Bollobás, 1979). Here the word "graph" is not used in its usual meaning of "diagram of a functional relationship", but as meaning "a collection of nodes and edges", in other words, a network. At the simplest level, the system's elements are reduced to graph nodes (also called vertices) and their interactions are reduced to edges connecting pairs of nodes (see Fig. 1). Edges can be either directed, specifying a source (starting point) and a target (endpoint), or non-directed. Directed edges are suitable for representing the flow of material from a substrate to a product in a reaction or the flow of information from a transcription factor to the gene whose transcription it regulates. Non-directed edges are used to represent mutual interactions, such as protein-protein binding. Graphs can be augmented by assigning various attributes to the nodes and edges; multi-partite graphs allow representation of different classes of node, and edges can be characterized by signs (positive for activation, negative for inhibition), confidence levels, strengths, or reaction speeds. Here I aim to show how graph representation and analysis can be used to gain biological insights through an understanding of the structure of cellular interaction networks.

## Graph concepts: from local to long-range

The nodes of a graph can be characterized by the number of edges that they have ( the number of other nodes to which they are adjacent). This property is called the **node degree**. In directed networks we distinguish the in-degree, the number of directed edges that point toward the node, and the out-degree, the number of directed edges that start at the node. Whereas node degrees characterize individual nodes, one can define a **degree distribution** to quantify the diversity of the whole network (see Fig. 1). The degree distribution *P(k)* gives the fraction of nodes that have degree *k* and is obtained by counting the number of nodes *N(k)* that have *k*=1, 2, 3… edges and dividing it by the total number of nodes *N*. The degree distribution of numerous networks, such as the World-wide web, Internet, human collaboration networks and metabolic networks, follows a well-defined functional form $P(k) = Ak^{-\gamma}$ called a power law . Here A is a constant that ensures that the *P(k)* values add up to one , and the **degree exponent** $\gamma$ is usually in the range 2<$\gamma$<3 (Albert and Barabási, 2002). This function indicates a high diversity of node degrees and that there is no typical node in the network that could be used to characterize the rest of the nodes (see Fig. 2). The absence of a typical degree (or typical scale) is why these networks are described as **"scale-free"**.

The cohesiveness of the neighborhood of a node *i* is usually quantified by the **clustering coefficient** $C_i$ , defined as the ratio between the number of edges linking nodes adjacent to *i* and the total possible number of edges among them (Watts and Strogatz, 1998). In other words, the clustering coefficient quantifies how close the local neighborhood of a node is to being part of a **clique**, a region of the graph (subgraph) where every node is connected to every other node. Various networks, including protein interaction and metabolic networks (Wagner and Fell, 2001; Yook et al., 2004) display a high average clustering coefficient, which indicates a high level of redundancy and cohesiveness. Averaging the clustering coefficients of nodes that have the same degree *k* gives the function *C(k)*, which characterizes the diversity of cohesiveness of local neighborhoods (see Fig. 1). Several measurements indicate a decreasing *C(k)* in metabolic networks (Ravasz et al., 2002) and protein interaction networks (Yook et al., 2004), following the relationship $C(k) = B/k^{\beta}$ where B is a constant and $\beta$ is between 1 and 2 . This suggests that low-degree nodes tend to belong to highly cohesive neighborhoods, whereas higher-degree nodes tend to have neighbors that are less connected to each other.

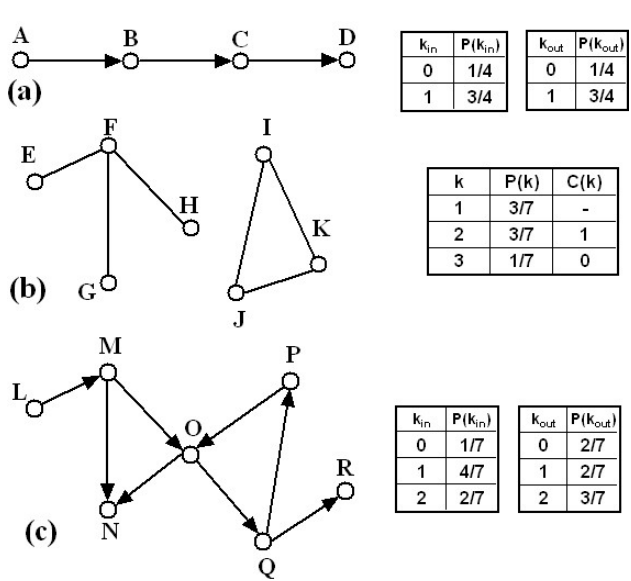

Figure 1. Graph representation and graph analysis reveals regulatory patterns of cellular networks. The number of interactions a component participates in is quantified by its (in/out) degree, for example node O has both in-degree and out-degree 2. The clustering coefficient characterizes the cohesiveness of the neighborhood of a node, for example the clustering coefficient of I is 1, indicating that it is part of a three-node clique. The graph distance between two nodes is defined as the number of edges in the shortest path between them. For example, the distance between nodes P and O is 1, and the distance between nodes O and P is 2 (along the OQP path). The degree distribution, $P(k)$ ( $P(k_{in})$ and $P(k_{out})$ in directed networks) quantifies the fraction of nodes with degree k, while the clustering-degree function $C(k)$ gives the average clustering coefficient of nodes with degree k.

(a) A linear pathway can be represented as a succession of directed edges connecting adjacent nodes. As there are no shortcuts or feedbacks in a linear pathway, and the distance between the starting and end node increases linearly with the number of nodes. The in and out-degree distribution indicates the existence of a source ($k_{in}=0$) and a sink ($k_{out}=0$) node.

(b) This undirected and disconnected graph is composed of two connected components (EFGH and IJK), has a range of degrees from 1 to 3 and a range of clustering coefficients from 0 (for F) to 1 (for I, J and). The connected component IJK is also a clique (completely connected subgraph) of three nodes.

(c) This directed graph contains a feed-forward loop (MON) and a feedback loop (POQ), which is also the largest strongly connected component of the graph. The in-component of this graph contains L and M, while its out-component consists of the sink nodes N and R. The source node L can reach every other node in the network.

Two nodes of a graph are connected if a sequence of adjacent nodes**, a path**, links them (Bollobás, 1979). A path can thus signify a transformation route from a nutrient to an end-product in a metabolic network, or a chain of post-translational reactions from the sensing of a signal to its intended target in a signal transduction network. The **graph distance** (also called path length) between two nodes is defined as the number of edges along the shortest path connecting them. If edges are characterized by the speed or efficiency of information propagation along them, the concept can be extended to signify, for example, the path with shortest delay (Dijkstra, 1959). In most networks observed, there is a relatively short path between any two nodes, and its length is on the order of the logarithm of the network size (Albert and Barabási, 2002; Newman, 2003b). This "**small world**" property appears to characterize most complex networks, including metabolic and protein interaction networks. If a path connects each pair of nodes, the graph is said to be **connected;** if this is not the case one can find **connected components,** graph regions (subgraphs) that are connected (see Fig. 1).

The connectivity structure of directed graphs presents special features, because the path between two nodes *i* and *j* can be different when going from *i* to *j* or vice versa (see Fig. 1). Directed graphs can have one or several **strongly connected components**, subgraph whose nodes are connected in both directions; **in-components**, which are connected to the nodes in the strongly connected component but not vice versa; and **out-components,** that can be reached from the strongly connected component but not vice versa. It is important to note that this topological classification reflects **functional separation** in signal transduction and metabolic networks. For example the regulatory architecture of a mammalian cell (Ma'ayan et al., 2004) has ligand-receptor binding as the in-component, a central signaling network as the strongly connected component and the transcription of target genes and phenotypic changes as part of the out-component.

The **source nodes** of directed cellular networks (the nodes that only have outgoing edges) can be regarded as corresponding to their inputs. For example, the substrates consumed from the environment (and not synthesized by the cell) constitute the inputs of a metabolic network, extracellular ligands or their receptors are the sources of signal transduction networks (Ma'ayan et al., 2005), and environmentally (but not transcriptionally ) regulated transcription factors constitute the sources of transcriptional networks. Following the paths starting from each source node will reveal a subgraph (termed **origon** in the context of transcriptional networks, (Balázsi et al., 2005) whose nodes can be potentially influenced by functional changes in the source node.

## Graph models

To understand how the above-defined graph measures reflect the organization of the underlying networks, we consider three representative graph families that have had a significant impact on network research (Albert and Barabási, 2002; Barabási and Oltvai, 2004; Newman, 2003b). A **linear pathway** has a well-defined source, a chain of intermediary nodes, and a sink (end) node. The clustering coefficient of each node is zero, because there are no edges among first neighbors. Both the maximum and average path length increase linearly with the number of nodes and are long for pathways that have many nodes (see Fig.1a). This type of graph has been widely used as a model of an isolated signal transduction pathway.

**Random graphs**, constructed by randomly connecting a given number *N* of nodes by *E* edges, reflect the (statistically) expected properties of a network of this size (Bollobás, 1985). They have a bell-shaped degree distribution (see Fig. 2), indicating that the majority of nodes have a degree close to the average degree <*k*>. The average clustering coefficient of a random graph $\langle C \rangle$ equals $\langle k \rangle / N$, thus is very small for large *N* (Albert and Barabási, 2002). Also, the *C(k)* function is a constant, indicating that the size of a local neighborhood does not influence its chance of being a clique. Thus random graphs are statistically homogeneous, because very small and very large node degrees and clustering coefficients are very rare. The average distance between nodes of a random graph depends logarithmically on the number of nodes, which results in very short characteristic paths (Bollobás, 1985).

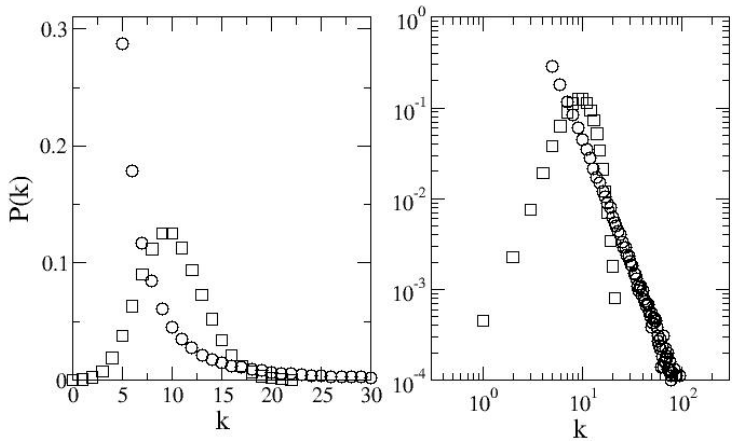

Figure 2: Comparison between the degree distribution of scale-free networks (circles) and random graphs (squares) having the same number of nodes and edges. For clarity the same two distributions are plotted both on a linear (left) and logarithmic (right) scale. The bell shaped degree distribution of random graphs peaks at the average degree and decreases fast for both smaller and larger degrees, indicating that these graphs are statistically homogeneous. In contrast, the degree distribution of the scale-free network follows the power law $P(k) = Ak^{-3}$, which appears as a straight line on a logarithmic plot. The continuously decreasing degree distribution indicates that low-degree nodes have the highest frequencies; however there is a broad degree range with non-zero abundance of very highly connected nodes (hubs) as well. Note that the nodes in a scale-free network do not fall into two separable classes corresponding to low-degree nodes and hubs, but every degree between these two limits appears with a frequency given by $P(k)$.

**Scale-free random graphs** are constructed such that they conform to a prescribed scale-free degree distribution but are random in all other aspects. Similar to scrambled but degree-preserving versions of real networks, these graphs serve as a much better suited null model of biological networks than random graphs, and indeed they have been used as comparison in identifying the significant interaction motifs of cellular networks (Milo et al., 2002; Shen-Orr et al., 2002) . Scale-free random graphs have even smaller path-lengths than random graphs (Cohen, 2003), and they are similar to random graphs in terms of their local cohesiveness (Newman, 2003a).

**Growing network models** strive to arrive at realistic topologies by describing network assembly and evolution. The simplest such model, (Barabási and Albert, 1999), incorporates two mechanisms: growth, (i.e. an increase in the number of nodes and edges over time) and preferential attachment, (i.e. an increased chance of high-degree nodes in acquiring new edges). Networks generated in this way have a power-law degree distribution $P(k) = Ak^{-3}$ (see Fig. 2), thus they can describe the higher end of the observed degree exponent range. Similarly to random graphs and scale-free random graphs, the average clustering coefficient in this model is small, and the clustering-degree function $C(k)$ is constant (Ravasz et al., 2002). The average path length is slightly smaller than that in comparable random graphs (Bollobás, 2003). The numerous improvements to this generic model include the incorporation of network evolution constraints and the identification of system-specific mechanisms responsible for preferential attachment (Albert and Barabási, 2002). Another growing network model, proposed by Ravasz et al. (2002), grows by iterative network duplication and integration to its original core. This growth algorithm leads to well-defined values for the node degree (for example, k=4, 5,

20, 84 when starting from a five-node seed) and clustering coefficient. The degree distribution can be approximated by a power law in which the exponent equals $\gamma = 1 + \log(n)/\log(n-1)$, where *n* is the size of the seed graph. Thus this model generates degree exponents in the neighborhood of 2, which is closer to the observed values than the degree exponent of the Barabási and Albert model. In contrast to all previous models, and in agreement with protein interaction and metabolic networks, the average clustering coefficient of the Ravasz et al. network does not depend on the number of nodes, and the clustering-degree function is heterogeneous, $C(k) \cong 1/k$, thus agrees with the lower range of observed clustering-degree exponent β.

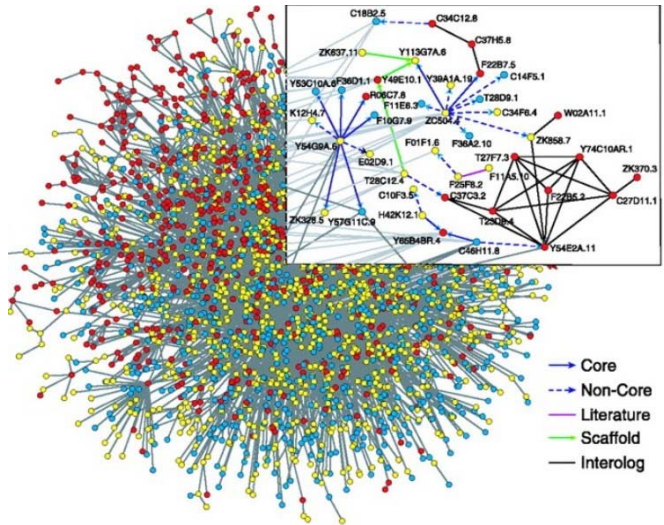

Figure 3: C. elegans protein interaction network. The nodes are colored according to their phylogenic class: ancient (red), multicellular (yellow) and worm (blue). The inset highlights a small part of the network. Figure reproduced from (Li et al., 2004b).

## From general to specific: properties of select cellular networks

During the last decade, genomics, transcriptomics and proteomics have produced an incredible quantity of molecular interaction data, contributing to maps of specific cellular networks (Burge, 2001; Caron et al., 2001; Pandey and Mann, 2000). In **protein interaction graphs,** the nodes are proteins, and two nodes are connected by a nondirected edge if the two proteins bind (see Fig. 3). Protein-protein interaction maps have been constructed for a variety of organisms, including viruses (McCraith et al., 2000), prokaryotes such as H. pylori (Rain et al., 2001) and eukaryotes such as S. cerevisiae (Gavin et al., 2002; Ho et al., 2002; Ito et al., 2001; Uetz et al., 2000), C. elegans (Li et al., 2004b) and D. melanogaster (Giot et al., 2003).

The current versions of protein interaction maps are, by necessity, incomplete, and also suffer from a high rate of false positives. Despite these drawbacks, there is an emerging consensus in the topological features of the maps of different organisms (see Fig. 4). For example, all protein interaction networks have a giant connected component and the distances on this component are close to the small-world limit given by random graphs (Giot et al., 2003; Yook et al., 2004). This finding suggests an illustration of pleiotropy, since perturbations of a single gene or protein can propagate through the network, and have seemingly unrelated effects. The degree distribution of the yeast protein interaction network is approximately scale-free (see Fig. 4). The Drosophila protein network exhibits a lower-than-expected fraction of proteins with more than 50 interacting partners; this deviation is suspected to be caused by incomplete coverage and could change as more interactions are discovered, as was the case for the yeast protein interaction network (Giot et al., 2003; Jeong et al., 2001; Yook et al., 2004). The heterogeneous clustering-degree function $C(k) = B/k^{\beta}$, where the exponent β is around 2 (Yook et al., 2004), and the inverse correlation between the degree of two interacting proteins (Maslov and Sneppen, 2002) indicate that the neighborhood of highly connected proteins tends to be sparser than the neighborhood of less connected proteins.

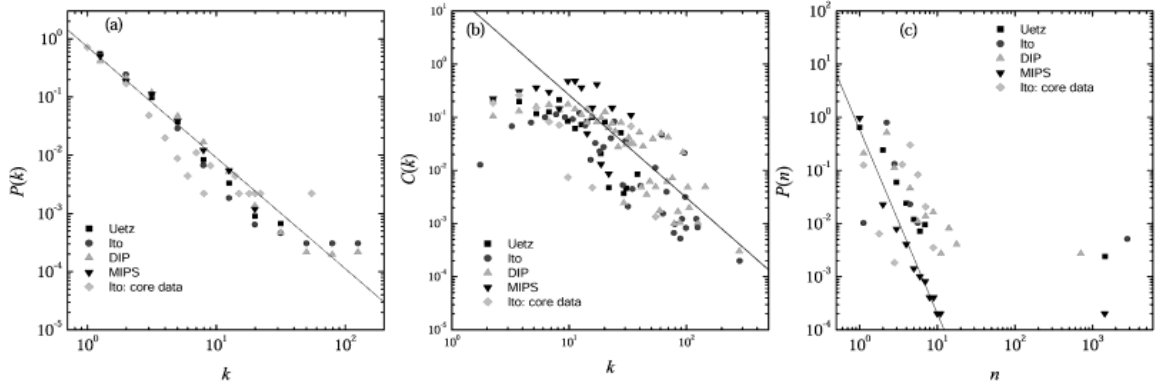

Figure 4. Topological properties of the yeast protein interaction network constructed from four different databases. a) Degree distribution. The solid line corresponds to a power law with exponent γ=2.5. b) Clustering coefficient – degree function. The solid line corresponds to the function $C(k) = B/k^2$. C) The size distribution of connected components. All networks have a giant connected component of more than 1000 nodes (on the right) and a number of small isolated clusters. Figure reproduced from Yook et al. 2004.

Arguably the most detailed representation of a network of reactions such as **the metabolic network** is a directed and weighted tri-partite graph, whose three types of node are metabolites, reactions and enzymes, and two types of edge represent mass flow and catalytic regulation, respectively (see Fig. 5). Mass flow edges connect reactants to reactions and reactions to products, and are marked by the stoichiometric coefficients of the metabolites (Feinberg, 1980; Lemke et al., 2004); enzymes catalyzing the reactions are represented as connected by regulatory edges to the nodes signifying the reaction (Jeong et al., 2000). Several simplified representations have also been studied - for example, the substrate graph, whose nodes are reactants, joined by an edge if they occur in the same chemical reaction (Wagner and Fell, 2001) or the reaction graph, whose nodes are reactions, connected if they share at least one metabolite.

All metabolic network representations indicate an approximately scale-free (Jeong et al., 2000; Tanaka, 2005; Wagner and Fell, 2001) or at least broad-tailed (Arita, 2004) metabolite degree distribution (see Fig. 6). The degree distribution of enzymes is strongly peaked, indicating that enzymes catalyzing several reactions are rare (Jeong et al., 2000). The variability of metabolite degrees can be accounted for if they are functionally separated into high-degree carriers and low-degree metabolites unique to separate reaction modules (such as catabolism or amino acid biosynthesis) (Tanaka, 2005) ; however, such a picture does not seem to explain the frequency of intermediate degrees. The clustering-degree function follows the relationship $C(k) \cong 1/k$.

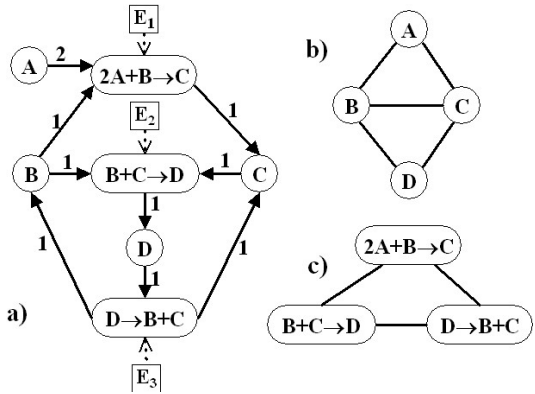

Figure 5. Three possible representations of a reaction network with three enzyme-catalized reactions and four reactants. The most detailed picture, a), includes three types of node: reactants (circles), reactions (rectangles) and enzymes (squares) and two types of edge corresponding to mass flow (solid lines) or catalysis (dashed lines). The edges are marked by the stochiometric coefficients of the reactants. b) In the metabolite network all reactants that participate in the same reaction are connected, thus the network is composed of a set of completely connected subgraphs (triangles in this case). c) In the reaction network two reactions are connected if they share a reactant. A similar graph can be constructed for the enzymes as well.

The substrate and reaction graphs indicate a remarkably small and organism-independent average distance among metabolites and reactions (Jeong et al., 2000; Wagner and Fell, 2001). If the preferred directionality of the reactions is known and is taken into account, only the largest strongly connected component (whose nodes can reach each other in both directions) has well-defined average path length. While this average path length is still small in all the organisms studied, the strongly connected component itself contains less than 50% of the nodes (Ma and Zeng, 2003). An alternative representation of the E. coli metabolic network defines edges among metabolites as structural changes that take convert the source metabolite into the target metabolite (Arita, 2004). As separate reactions can involve the same structural change in a metabolite, this alternative representation has less than half as many edges than the metabolite graph defined by (Jeong et al., 2000), and consequently it yields twice as high average metabolite distances.

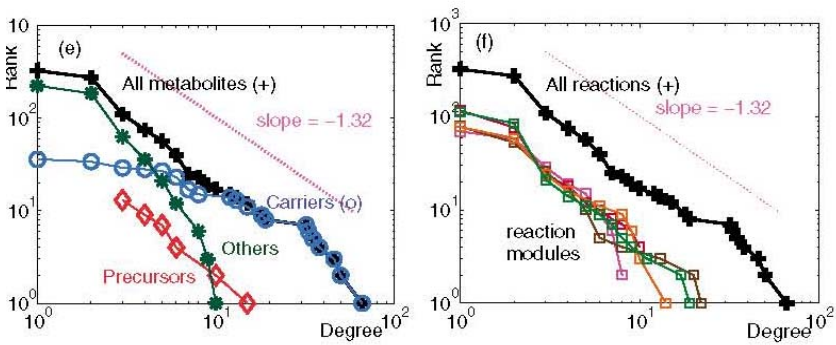

Figure 6. Rank (cumulative distribution) of metabolite node degree (left panel) and reaction node degree (right panel) for metabolic networks of *H. pylori*. The straight lines correspond to a power-law degree distribution with exponent $\gamma$=slope+1=2.32. The figure illustrates that functionally different metabolites tend to cover different ranges of the degree spectrum. Reproduced from (Tanaka, 2005).

It is now possible to identify the set of target genes for each transcription factor encoded by a cell, and **transcriptional regulatory maps** have been constructed for E. coli (Shen-Orr et al., 2002) and S. cerevisiae (Guelzim et al., 2002; Lee et al., 2002; Luscombe et al., 2004). The full representation of such a network has two types of node, which correspond to transcription factors and the mRNAs of the target genes, and two types of directed edge, which correspond to transcriptional regulation and translation (Lee et al., 2002). For simplicity, transcription factors are often combined with the genes encoding them; thus all nodes correspond to genes (see Fig. 7). The nodes representing target genes that do not encode transcription factors become sinks while non-transcriptionally regulated transcription factors correspond to sources.

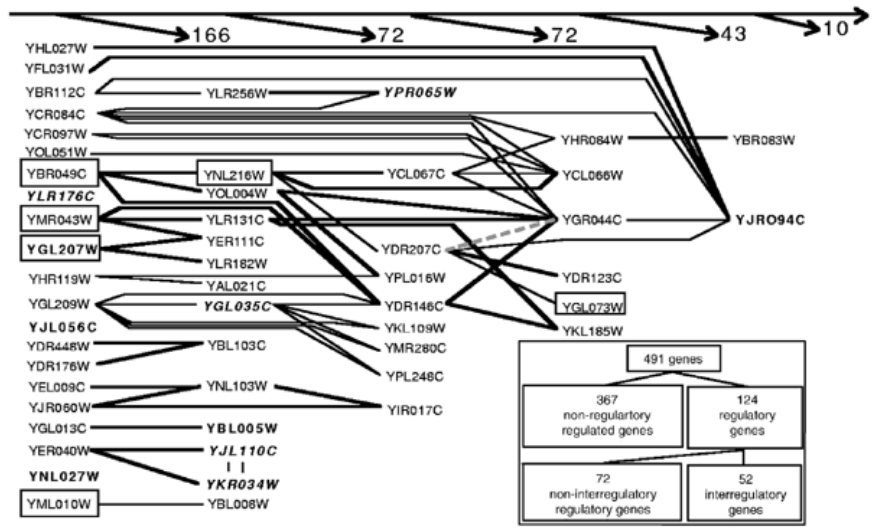

Figure 7. Interactions among 52 inter-regulatory genes in the transcriptional regulatory network of S. cerevisiae. The gene names are arranged in such a way that left to right illustrates downstream causality. The non-regulatory genes regulated by each column of regulatory genes are shown on the top arrow. Bold type indicates self-activation, bold italics indicates self-inhibition, and borders indicate essential genes. Reproduced with the permission of the Nature Publishing Group from (Guelzim et al., 2002).

Both prokaryotic and eukaryotic transcription networks exhibit an approximately scale-free out-degree distribution, signifying the potential of transcription factors to regulate a multitude of target genes. The in-degree distribution is a more restricted exponential function, illustrating that combinatorial regulation by several transcription factors is observed less than regulation of several targets by the same transcription factor (see Fig. 8). Neither the E. coli nor the yeast transcription network have strongly connected components, indicating a unidirectional, feed-forward type regulation mode. The subgraphs found by following the paths that start from non-transcriptionally regulated genes have relatively little overlap (Balázsi et al., 2005), reflecting that distinct environmental signals tend to initiate distinct transcriptional responses. The source – sink distances are small in both networks, and the longest regulatory chain has only four (in E. coli) respectively five (in S. cerevisiae) edges (see Fig. 8).

Elucidation of the mechanisms that connect extracellular signal inputs to the control of transcription factors was until recently restricted to small-scale biochemical, genetic and pharmacological intervention techniques. **Signal transduction pathways** have traditionally been viewed as linear chains of biochemical reactions and protein-protein interactions, starting from signal sensor molecules and reaching intracellular targets, however the increasingly recognized abundance of components shared by several pathways indicates that an interconnected signaling network exists[1]. The largest reconstructed signal transduction network contains 1259 interactions among 545 cellular components of the hippocampal CA1 neuron (Ma'ayan et al., 2005), based on more than 1200 articles in the experimental literature. This network exhibits an impressive interconnectivity: its strongly connected component (the central signaling network) includes 60% of the nodes, and the subgraphs that start from various ligand-occupied receptors reach most of the network within 15 steps. The average input-output path-length is near 4, suggesting the possibility of very rapid response to signaling inputs. Both the in- and out-degree distribution of this network is consistent with a power-law with an exponent around 2, the highest degree nodes including the four major protein kinases (MAPK, CaMKII, PKA and PKC).

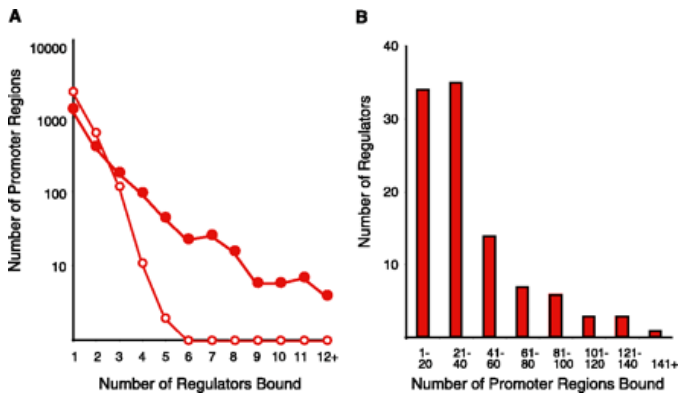

Figure 8. Genome-wide distribution of transcriptional regulators in S. cerevisiae. A. Full symbols represent the number of transcription factors bound per promoter region (corresponding to the in-degree of the regulated gene). Open symbols stand for a the in-degree distribution of a comparable randomized network. B. Distribution of the number of promoter regions bound per regulator (i.e. the out-degree distribution of transcription factors). Figure reproduced from (Lee et al., 2002).

In addition to the networks whose edges signify biological interactions, several **functional association networks** based on gene co-expression (Stuart et al., 2003; Valencia and Pazos, 2002), gene fusion or co-occurrence (von Mering et al., 2002) or genetic interactions (Tong et al., 2004) have been constructed. For example, synthetic lethal interactions, introduced between pairs of genes whose combined knock-out causes cell death, indicate that these genes buffer for one another (see Fig. 9). A recent study by Tong et al. (2004) shows that the yeast genetic interaction network has small world and scale free properties, having a small average path length, dense local neighborhoods, and an approximately power-law degree distribution. The overlap between the yeast protein interaction and genetic interaction network is extremely small, which is expected since genetic interactions reflect a complex functional compensatory relationship and not a physical interaction (see Fig. 9). Indeed, the gene relationships that do overlap with genetic interactions are: having the same mutant phenotype, encoding proteins with the same subcellular localization or encoding proteins within the same protein complex.

---

[1] We note here that despite the separate nomenclature there is a significant overlap between protein interaction networks, metabolic networks and signal transduction networks.

## Biological interpretation of graph properties
The architectural features of molecular interaction networks are shared to a large degree by other complex systems ranging from technological to social networks. While this universality is intriguing and allows us to apply graph theory to biological networks, we need to focus on the interpretation of graph properties in light of the functional and evolutionary constraints of cellular networks.

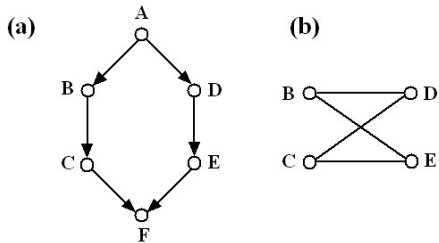

Figure 9. Connections between pathway redundancy and synthetic lethal interactions. (a) Consider a hypothetical cellular network module that receives exogenous signals through node A and whose sink node F determines the response to the signal (or the phenotype). There are two node-independent (redundant) pathways between nodes A and F that can compensate for each other in case of node disruptions. By defining synthetic lethal interactions as pairs of nodes whose loss causes the disconnection of nodes A and F, one would find graph (b). The two graphs present complementary and non-overlapping information.

**Hubs**: In a scale-free network small-degree nodes are the most abundant, but the frequency of high-degree nodes decreases relatively slowly. Thus, nodes that have degrees much higher than average, so-called hubs, exist. Because of the heterogeneity of scale-free networks, random node disruptions do not lead to a major loss of connectivity, but the loss of the hubs causes the breakdown of the network into isolated clusters (Albert and Barabási, 2002). The validity of these general conclusions for cellular networks can be verified by correlating the severity of a gene knockout with the number of interactions the gene's products participate in. Indeed, as much as 73% of the S. cerevisiae genes are non-essential, -i.e. the knockout has no phenotypic effects (Giaever et al., 2002). This confirms the cellular networks' robustness in the face of random disruptions. The likelihood that a gene is essential (lethal) or toxicity modulating (toxin sensitive) correlates with the number of interactions its protein product has (Jeong et al., 2001; Said et al., 2004). This indicates the cell is vulnerable to the loss of highly interactive hubs[2]. Among the most well-known examples of hub proteins is the tumor suppressor protein p53 that has an abundance of incoming edges, interactions regulating its conformational state (and thus its activity) and its rate of proteolytic degradation, and that also has a lot of outgoing edges in the genes whose transcription it activates. The tumor suppressor p53 is inactivated by a mutation in its gene in 50% of human tumors, in agreement with cellular networks' vulnerability to their most connected hubs (Vogelstein et al., 2000).

Given the importance of highly connected nodes, one can hypothesize that they are subject to severe selective and evolutionary constraints. (Hahn et al., 2004) have correlated the rate of evolution of yeast proteins with their degree in the protein interaction network, and the rate of evolution of E. coli enzymes with their degree in the core metabolic reaction graph constructed by (Wagner and Fell, 2001). Although they obtained statistically significant (albeit weak) negative correlation between yeast protein degree and evolution rate, no such correlation was evident in the E. coli enzyme network. The latter result has the caveat that the edges linking enzymes do not correspond to interactions; thus further studies are needed to gain a definitive answer.

---
[2] We note here that different network representations can lead to distinct sets of hubs and there is no rigid boundary between hub and non-hub genes or proteins.

**Modularity:** Cellular networks have long thought to be **modular**, composed of functionally separable subnetworks corresponding to specific biological functions (Hartwell et al., 1999). Since genome-wide interaction networks are highly connected, modules should not be understood as disconnected components but rather as components that have dense intra-component connectivity but sparse inter-component connectivity. Several methods have been proposed to identify functional modules on the basis of the physical location or function of network components (Rives and Galitski, 2003) or the topology of the interaction network (Giot et al., 2003; Girvan and Newman, 2002; Spirin and Mirny, 2003). The challenge is that modularity does not always mean clear-cut subnetworks linked in well-defined ways, but there is a high degree of overlap and cross-talk between modules (Han et al., 2004). As Ravasz et al. (2002) recently argued, a heterogeneous degree distribution, inverse correlation between degree and clustering coefficient (as seen in metabolic and protein interaction networks) and modularity taken together suggest **hierarchical modularity**, in which modules are made up of smaller and more cohesive modules, which themselves are made up of smaller and more cohesive modules etc.

**Motifs and cliques:** There is growing evidence suggesting that cellular networks contain conserved **interaction motifs,** small subgraphs that have well-defined topology. Interaction motifs such as autoregulation and feed-forward loops have a higher abundance in transcriptional regulatory networks than expected from randomly connected graphs with the same degree distribution (Balázsi et al., 2005; Shen-Orr et al., 2002). Protein interaction motifs such as short cycles and small completely connected subgraphs are both abundant (Giot et al., 2003) and evolutionarily conserved (Wuchty et al., 2003), partly because of their enrichment in protein complexes. Feedforward loops and triangles of scaffolding (protein) interactions are also abundant in signal transduction networks, which also contain a significant number of feedback loops, both positive and negative (Ma'ayan et al., 2005). Yeger-Lotem et al. identified frequent composite transcription/protein interaction motifs such as interacting transcription factors coregulating a gene or interacting proteins being coregulated by the same transcription factor (Yeger-Lotem et al., 2004). As (Zhang et al., 2005) have pointed out, the abundant motifs of integrated mRNA/protein networks are often signatures of higher-order network structures that correspond to biological phenomena (see Fig. 10). Conant and Wagner found that the abundant transcription factor motifs of E. coli and S. cerevisiae do not show common ancestry but are a result of repeated convergent evolution (Conant and Wagner, 2003). These findings, as well as studies of the dynamical repertoire of interaction motifs, suggest that these common motifs represent elements of optimal circuit design (Csete and Doyle, 2002; Ma'ayan et al., 2005; Mangan and Alon, 2003).

**Path redundancy:** Any response to a perturbation requires that information about the perturbation spreads within the network. Thus the short path lengths of metabolic, protein interaction and signal transduction networks (their **small world** property) (Jeong et al., 2001; Jeong et al., 2000; Ma'ayan et al., 2005) is a very important feature that ensures fast and efficient reaction to perturbations. Another very important global property related to paths is **path redundancy**, or the availability of multiple paths between a pair of nodes (Papin and Palsson, 2004). Either in the case of multiple flows from input to output, or contingencies in the case of perturbations in the preferred pathway, path redundancy enables the robust functioning of cellular networks by relying less on individual pathways and mediators. The frequency of nodes' participation in paths connecting other components can be quantified by their **betweenness centrality**, first defined in the context of social sciences (Wasserman and Faust, 1994). Node betweenness, adapted to the special conditions of signal transduction networks, can serve as an alternative measure for identifying important network hubs.

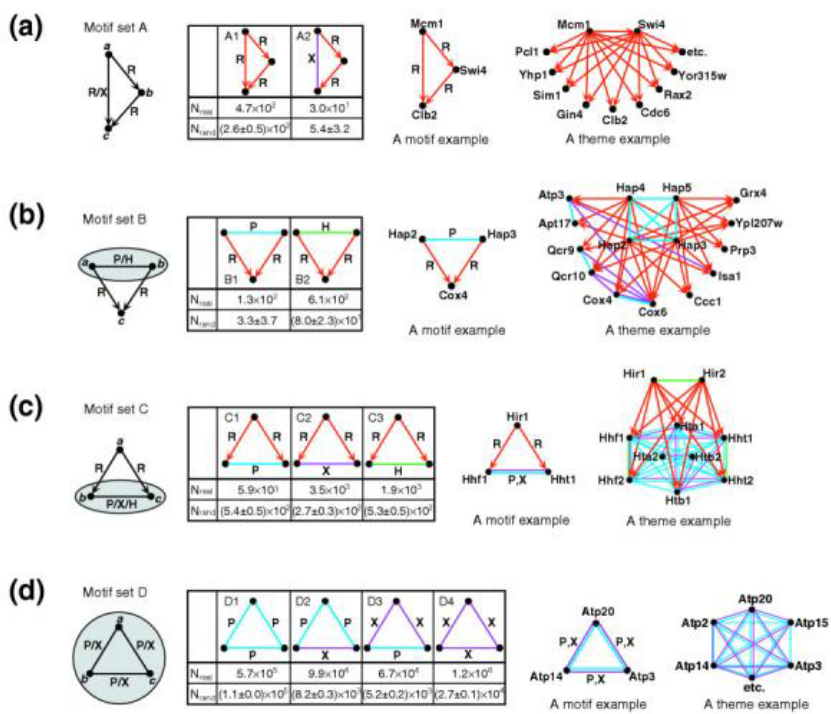

Figure 10. Network motifs and themes in the integrated S. cerevisiae network. Edges among genes denote transcriptional regulation (R), protein interaction (P), sequence homology (H), correlated expression (X) or synthetic lethal interactions (S). a) Motifs corresponding to the "feed-forward" theme are based on transcriptional feed-forward loops; b) motifs in the "co-pointing" theme consist of interacting transcription factors regulating the same target gene ; c) motifs corresponding to the "regulonic complex" theme include co-regulation of members of a protein complex; d) motifs in the "protein complex" theme represent interacting and coexpressed protein cliques. For a given motif, $N_{real}$ is the number of corresponding subgraphs in the real network, and $N_{rand}$ is the number of corresponding subgraphs in a randomized network. Figure reproduced from (Zhang et al., 2005).

## Network models specific to biological networks

The topology of cellular networks is shaped by dynamical processes on evolutionary time scales. These processes include gene or genome duplication and gain or loss of interactions due to mutations. Many researchers investigated whether the similar topological properties of biological networks and social or technological networks point towards shared growth principles, and whether variants of general growing network models apply to cellular networks as well. The most intriguing question is the degree to which natural selection, specific to biological systems, shapes the evolution of cellular network topologies.

Several growing network models based on random gene duplication and subsequent functional divergence have displayed good agreement with the topology of protein interaction networks (Kim et al., 2002; Pastor-Satorras et al., 2003; Vazquez A, 2003). However, estimates of gene duplication rate and the rate at which point mutations lead to the gain or loss of protein interactions indicate that point mutations are two orders of magnitude more frequent than gene duplications (Berg et al., 2004). Berg et al. (2004) propose a protein network evolution model based on edge dynamics and to a lesser extent, gene duplication, and find that it generates a topology similar to the yeast protein interaction network. It is interesting to note that both gene duplications and point mutations, specific biological processes, lead to a preferential increase of the degree of highly connected proteins, also confirmed by measurements (Eisenberg and Levanon, 2003; Wagner, 2003). Thus natural selection could affect the balance between interaction gain and loss in such a way that an effective preferential attachment is obtained. The modeling of the evolution of transcriptional, metabolic and signal transduction networks has added challenges due to their directed nature and to the complexity of the regulatory mechanisms involved, but rapid progress is expected in these fields as well (Light and Kraulis, 2004; Tanay et al., 2005).

## Beyond static properties

As illustrated in the specific examples presented in this review, graph representations of cellular networks and quantitative measures characterizing their topology can be extremely useful for gaining systems-level insights into cellular regulation. For example, the interconnected nature of cellular networks indicates that perturbations of a gene or protein could have seemingly unrelated effects (pleiotropy), a result that would seem counterintuitive in a reductionist framework. The graph framework allows us to discuss the cell's molecular makeup as a network of interacting constituents, and to shift the definition of "gene function" from an individual-based attribute to an attribute of the network (or network module) in which the gene participates (Fraser and Marcotte, 2004). Interaction motifs and themes can be exploited to predict individual interactions given sometimes-uncertain experimental evidence, or to give a short list of candidates for experimental testing (Albert and Albert, 2004; King et al., 2004; Wong et al., 2004).

It is important to realize that cellular interaction maps represent a network of possibilities, and not all edges are present and active at the same time or in a given cellular location *in vivo*. Indeed, superposing mRNA expression patterns and protein interaction information in S. cerevisiae, Han et al. (2004) identified a strong dynamical modularity mediated by two types of highly interactive proteins: party hubs, which interact with most of their partners simultaneously, and date hubs, which bind their different partners at different times or location. Similarly, Luscombe et al. (2004) and Balázsi et al. (2005) found that only subsets of the yeast and E. coli transcriptional networks are active in a given exogenous or endogenous condition, the former inducing only a few transcription factors with little crosstalk , while the latter activating connected clusters of transcription factors and many feedforward loops.

In addition, the diversity of metabolic fluxes (Almaas et al., 2004) and reaction rates/timescales (Papin et al., 2005) attest that only an integration of interaction and activity information will be able to give the correct dynamical picture of a cellular network (Levchenko, 2003; Ma'ayan et al., 2004). To move significantly beyond our present level of knowledge, new tools for quantifying concentrations, fluxes and interaction strengths, in both space and time, are needed. In the absence of comprehensive time-course datasets, dynamical reconstruction and analysis can be usually carried out for small networks only (Hoffmann et al., 2002; Lee et al., 2003; Tyson et al., 2001). The coupling of experimental data with mathematical modeling enables the identification of previously unknown regulatory mechanisms. For example, the (Hoffmann et al., 2002) model's prediction regarding the importance of particular IκB isoforms in feedback loops regulating NF-κB was experimentally verified, as were the dynamic profiles of β-catenin concentrations in the model of the WNT signaling module by (Lee et al., 2003).

The currently limited knowledge of kinetic parameters makes the construction of detailed kinetic models of complex biological networks next to impossible; however, there is hope that more coarse-grained models can also be successful. Indeed, there is increasing evidence of the crucial role of network topology in determining dynamical behavior and function, and of robustness to fluctuations in kinetic parameters (Albert and Othmer, 2003; Barkai and Leibler, 1997; Chaves, 2005; Li et al., 2004a; von Dassow et al., 2000). The topological properties of signal transduction subgraphs (pathways) seem to reflect the dynamics of response to those signals: the subgraphs corresponding to ligands that cause rapid, transient changes -such as glutamate or glycine- exhibit extensive pathway branching, while the signaling pathways of FasL or ephrin have much fewer branches (Ma'ayan et al., 2005). Constraint-based modeling of stochiometrically reconstructed metabolic and signaling networks can lead to verifiable predictions related to their input/output relationship and its changes in case of gene knockouts (Papin and Palsson, 2004; Papin et al., 2002). Taken together, network discovery and network analysis have the potential to form a self-reinforcing loop where theory and modeling leads to testable predictions that feed back into experimental discovery.

This review focused on the insights gained from analyzing the topology of cellular networks; for information on other important related topics such as computational methods of network inference and mathematical modeling of the dynamics of cellular networks I suggest the excellent review articles (Friedman, 2004; Longabaugh et al., 2005; Ma'ayan et al., 2004; Papin et al., 2005; Tyson et al., 2003). At a minimum, network representations have changed our view of what is functionally "downstream" (or "near") a cellular component, and can potentially lead to predictions of systems-level behavior that will be important for future biochemical and medical research (Cohen, 2002).

**Acknowledgements**: The author gratefully acknowledges the support of a Sloan Fellowship in Science and Engineering.